\documentstyle[preprint,tighten,aps,psfig]{revtex}

\begin{document}

%
\draft
\preprint{FERMILAB-PUB-96/449-E, D0PUB-96-15}
\title{ Search for Top Squark Pair Production in the Dielectron Channel}

%
\author{                                                                        
S.~Abachi,$^{14}$                                                               
B.~Abbott,$^{28}$                                                               
M.~Abolins,$^{25}$                                                              
B.S.~Acharya,$^{43}$                                                            
I.~Adam,$^{12}$                                                                 
D.L.~Adams,$^{37}$                                                              
M.~Adams,$^{17}$                                                                
S.~Ahn,$^{14}$                                                                  
H.~Aihara,$^{22}$                                                               
G.~\'{A}lvarez,$^{18}$
% ...
\\                                                                              
\vskip 0.50cm                                                                   
\centerline{(D\O\ Collaboration)}                                               
\vskip 0.50cm                                                                   
}                                                                               
\address{                                                                       
\centerline{$^{1}$Universidad de los Andes, Bogot\'{a}, Colombia}               
\centerline{$^{2}$University of Arizona, Tucson, Arizona 85721}                 
\centerline{$^{3}$Boston University, Boston, Massachusetts 02215}               
\centerline{$^{4}$Brookhaven National Laboratory, Upton, New York 11973}        
\centerline{$^{5}$Brown University, Providence, Rhode Island 02912}             
\centerline{$^{6}$Universidad de Buenos Aires, Buenos Aires, Argentina}         
\centerline{$^{7}$University of California, Davis, California 95616}            
\centerline{$^{8}$University of California, Irvine, California 92717}           
\centerline{$^{9}$University of California, Riverside, California 92521}        
\centerline{$^{10}$LAFEX, Centro Brasileiro de Pesquisas F{\'\i}sicas,          
                  Rio de Janeiro, Brazil}                                       
\centerline{$^{11}$CINVESTAV, Mexico City, Mexico}                              
\centerline{$^{12}$Columbia University, New York, New York 10027}               
\centerline{$^{13}$Delhi University, Delhi, India 110007}                       
\centerline{$^{14}$Fermi National Accelerator Laboratory, Batavia,              
                   Illinois 60510}                                              
\centerline{$^{15}$Florida State University, Tallahassee, Florida 32306}        
\centerline{$^{16}$University of Hawaii, Honolulu, Hawaii 96822}                
\centerline{$^{17}$University of Illinois at Chicago, Chicago, Illinois 60607}  
\centerline{$^{18}$Indiana University, Bloomington, Indiana 47405}              
\centerline{$^{19}$Iowa State University, Ames, Iowa 50011}                     
\centerline{$^{20}$Korea University, Seoul, Korea}                              
\centerline{$^{21}$Kyungsung University, Pusan, Korea}                          
\centerline{$^{22}$Lawrence Berkeley National Laboratory and University of      
                   California, Berkeley, California 94720}                      
\centerline{$^{23}$University of Maryland, College Park, Maryland 20742}        
\centerline{$^{24}$University of Michigan, Ann Arbor, Michigan 48109}           
\centerline{$^{25}$Michigan State University, East Lansing, Michigan 48824}     
\centerline{$^{26}$Moscow State University, Moscow, Russia}                     
\centerline{$^{27}$University of Nebraska, Lincoln, Nebraska 68588}             
\centerline{$^{28}$New York University, New York, New York 10003}               
\centerline{$^{29}$Northeastern University, Boston, Massachusetts 02115}        
\centerline{$^{30}$Northern Illinois University, DeKalb, Illinois 60115}        
\centerline{$^{31}$Northwestern University, Evanston, Illinois 60208}           
\centerline{$^{32}$University of Notre Dame, Notre Dame, Indiana 46556}         
\centerline{$^{33}$University of Oklahoma, Norman, Oklahoma 73019}              
\centerline{$^{34}$University of Panjab, Chandigarh 16-00-14, India}            
\centerline{$^{35}$Institute for High Energy Physics, 142-284 Protvino, Russia} 
\centerline{$^{36}$Purdue University, West Lafayette, Indiana 47907}            
\centerline{$^{37}$Rice University, Houston, Texas 77005}                       
\centerline{$^{38}$Universidade Estadual do Rio de Janeiro, Brazil}             
\centerline{$^{39}$University of Rochester, Rochester, New York 14627}          
\centerline{$^{40}$CEA, DAPNIA/Service de Physique des Particules, CE-SACLAY,   
                   France}                                                      
\centerline{$^{41}$Seoul National University, Seoul, Korea}                     
\centerline{$^{42}$State University of New York, Stony Brook, New York 11794}   
\centerline{$^{43}$Tata Institute of Fundamental Research,                      
                   Colaba, Bombay 400005, India}                                
\centerline{$^{44}$University of Texas, Arlington, Texas 76019}                 
\centerline{$^{45}$Texas A\&M University, College Station, Texas 77843}         
}                                                                               
\date{\today}

\maketitle

\begin{abstract}
This report describes the first search for top squark pair production in the
channel $\widetilde t_1\overline{\widetilde t}_1 \to
b\overline{b} \widetilde{\chi}^{+}_1 \widetilde{\chi}^{-}_1 \to e e
$+jets+\mbox{${\hbox{$E$\kern-0.6em\lower-.1ex\hbox{/}}}_T$}\ 
using $74.9 \pm 8.9$ pb$^{-1}$\
of data collected using the D\O\ detector.
A 95\% confidence level upper limit on  $\sigma\cdot B$
is presented.  The limit is above the theoretical expectation for
$\sigma\cdot B$ for this process, but does show the sensitivity of the current
D\O\ data set to a particular topology for new physics.
\end{abstract}
\pacs{PACS numbers: 14.80.Ly, 13.85.Rm}
%
%
%
                                                                                
Supersymmetry (SUSY) is a fundamental space-time symmetry relating bosons and
fermions \cite{susy}.
Supersymmetric extensions to the Standard Model (SM) feature undiscovered
superpartners for every SM particle --- for example, there is a scalar
quark (squark) for
each of the two degrees of freedom for the spin 1/2 quarks.  In most SUSY
models, the masses of the squarks are approximately degenerate except for 
those of the top squarks.  Due to large top family Yukawa interactions,
the lighter top squark mass eigenstate ($\widetilde t_1$) can 
have a much lower mass than the other squarks\cite{Howie}.
Top squarks will be pair produced at the Tevatron; each will then decay into
the lightest chargino
$\widetilde{\chi}^\pm_1$\ and a $b$ quark if that decay is
kinematically allowed.
If $m_{\widetilde{\chi}^\pm_1}$\ is greater than the mass of the top squark,
the decay will
proceed through slepton channels.  If the sleptons are also heavier than
the top squark, it will decay to a charm quark and the
lightest supersymmetric particle (LSP or $\widetilde{\chi}^0_1$)
with a 100\% branching fraction.
                                                                                
The $\widetilde{\chi}^\pm_1$\ decays to $l\nu_l\widetilde{\chi}^0_1$ or
$q{\overline q}^\prime \widetilde{\chi}^0_1$.  Under $R$-parity conservation,
the lightest neutralino is assumed to be stable and escapes detection,
resulting in missing transverse energy 
\mbox{${\hbox{$E$\kern-0.6em\lower-.1ex\hbox{/}}}_T$}.
Thus, top squarks, pair-produced at the Tevatron, result in final states
similar to those of top quarks.  However, as the decay of the chargino is
to three particles,
the decay products tend to be softer than those of the $W$ boson.
Since the $\widetilde{\chi}^\pm_1$\ decay is
almost always dominated by virtual $W$ exchange,
the branching fractions are expected to be be very close to $W$ boson 
leptonic and hadronic decay branching fractions\cite{Howie}.
                                                                                
The results of a search for
$\widetilde t_1\to c\widetilde{\chi}^0_1$\ have been published by the D\O\
Collaboration \cite{stopprl}.
Model independent lower limits on the masses of the top squark and lightest
chargino have been set using the measured width of the $Z$ boson
and are approximately 45 GeV/c$^2$ \cite{zlimits}.
Within the framework of the Minimal Supersymmetric
Standard Model (MSSM) \cite{mssm}, the current limits on the pair production
of charginos (which depend on the assumed value of the common scalar mass
$m_0$) from LEP at
$\sqrt{s}=$ 130, 136, and 161 GeV \cite{opal,lots_of_refs},
lead to $m_{\widetilde{\chi}^\pm_1} > 62.0$ -- 78.5 GeV/c$^2$
at the 95\% CL \cite{opal}.
The analysis described below is independent of the MSSM and supergravity
\cite{sugra} frameworks,
instead depending only on the masses of the top squark, the
lightest chargino, and the lightest neutralino, and on the branching fractions
of the chargino decay.
                                                                                
Previous phenomological studies have considered final states with a
single lepton + jets +
\mbox{${\hbox{$E$\kern-0.6em\lower-.1ex\hbox{/}}}_T$}\ and two leptons +
jets + \mbox{${\hbox{$E$\kern-0.6em\lower-.1ex\hbox{/}}}_T$} \cite{Howie}.
These studies, which used {\small ISAJET} \cite{isajet} events smeared by
typical detector resolutions, indicated that the single
lepton channel cannot be studied at the Tevatron without excellent
$b$-tagging capability due to the enormous background from $W$ boson
production.
However, they did indicate that analysis of the dilepton channels
($ee$, $e\mu$ and $\mu\mu$) could lead to a limit on the mass of
(or discovery of) the top squark at the Tevatron using the current data set.
                                                                                
This report
describes the first search for the decay $\widetilde t_1\to
b\widetilde{\chi}^\pm_1$\ in the
channel $\widetilde t_1\overline{\widetilde t}_1 \to  e e
$+ jets + \mbox{${\hbox{$E$\kern-0.6em\lower-.1ex\hbox{/}}}_T$}\ 
using $74.9 \pm 8.9$ pb$^{-1}$\ of data.
The data were collected at the Fermilab Tevatron at
$\sqrt{s} = 1.8$ TeV during 1994--1995.
The D\O\ detector and data collection system are described in detail in
Ref. \cite{d0nim}.  The detector consisted of three major subsystems:
a uranium-liquid argon calorimeter, central tracking detectors (with no
central magnetic field), and a muon spectrometer.  Electrons were identified
by their longitudinal and transverse shower profiles in the calorimeter and
were required to have a matching track in the central tracking chambers.
In this analysis, they were restricted to have pseudorapidity
$|{\eta}| < 2.5$ and to be
isolated from other energy depositions in the event.
Jets were reconstructed using a cone algorithm of radius
${\cal{R}} = \sqrt{(\Delta\phi)^2 +(\Delta\eta)^2} = 0.7$ with
$|{\eta}| < 4.0$.  The
\mbox{${\hbox{$E$\kern-0.6em\lower-.1ex\hbox{/}}}_T$}\ was determined from
energy deposition in the calorimeter for $|{\eta}| < 4.5$.
                                                                                
The acceptance for top squark events was calculated for a range of
top squark and chargino masses using the {\small ISAJET} event generator
and a detector simulation based on the {\small GEANT} program \cite{geant}.
Samples were generated with top squark masses between 55 and
75 GeV/c$^2$\ with $m_{\widetilde{\chi}^\pm_1}$\ between 47 and 68 GeV/c$^2$,
depending on $m_{\widetilde t_1}$.
The mass of the lightest neutralino was set to the supergravity-motivated
value ${1\over{2}}m_{\widetilde{\chi}^\pm_1}$.
                                                                                
\begin{figure}\vbox{
\centerline{\psfig{figure=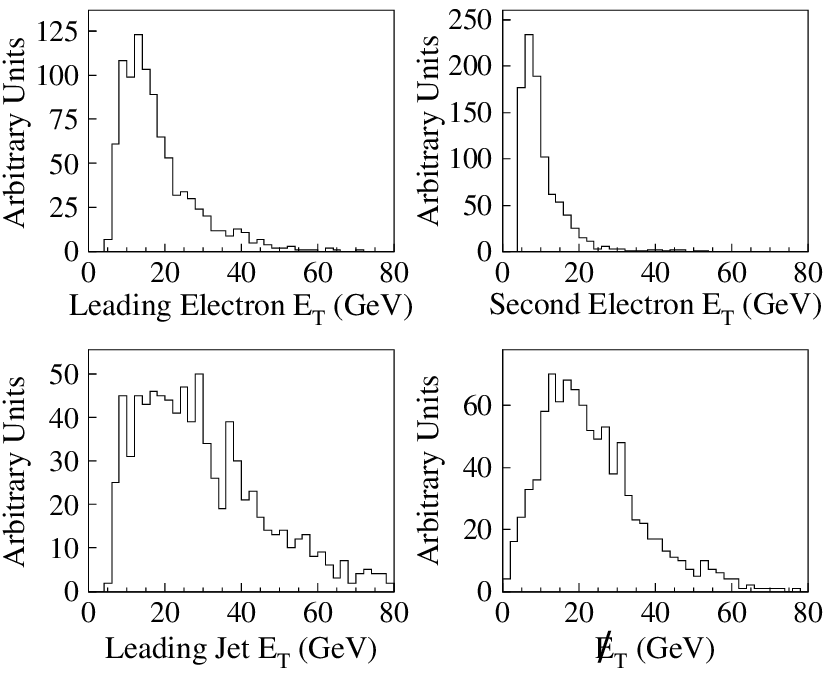,width=8.5cm}}
\caption{Kinematic distributions for
($m_{\widetilde t_1}$,$m_{\widetilde{\chi}^\pm_1}$) = (65,47)
GeV/c$^2$. }
\label{fig:kine}
}
\end{figure}

The signature for $\widetilde t_1 \overline{\widetilde t}_1 \to
b\overline{b} \widetilde{\chi}^{+}_1 \widetilde{\chi}^{-}_1$\ 
is two electrons, one or more jets, and
\mbox{${\hbox{$E$\kern-0.6em\lower-.1ex\hbox{/}}}_T$}.
Kinematic distributions for
($m_{\widetilde t_1}$,$m_{\widetilde{\chi}^\pm_1}$) = (65,47)
GeV/c$^2$\
are shown in Fig. \ref{fig:kine}.
This analysis was restricted to events selected using a trigger which
required one electromagetic cluster with transverse energy $E_T^e > 15$ GeV,
one jet with $E_T^j > 10$ GeV, and
\mbox{${\hbox{$E$\kern-0.6em\lower-.1ex\hbox{/}}}_T$}\ $>$ 14 GeV.
Other kinematic quantities used to discriminate against background are
the invariant mass of the two electrons $m_{ee}$ and 
$E_T^{\text{sum}} = E_T^{e1} + E_T^{e2} 
+ {\mbox{${\hbox{$E$\kern-0.6em\lower-.1ex\hbox{/}}}_T$}}$ 
(defined in Ref. \cite{Howie} as bigness).
                                                                                
\begin{figure}\vbox{
\centerline{\psfig{figure=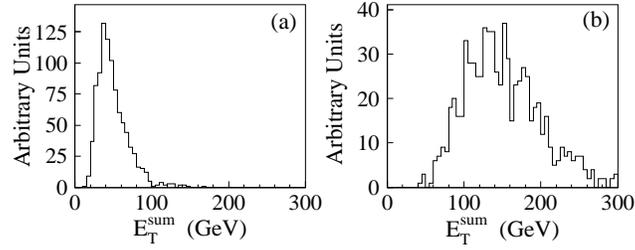,width=8.5cm}}
\caption{Distributions of $E_T^{\text{sum}}$\ for (a)
($m_{\widetilde t_1}$,$m_{\widetilde{\chi}^\pm_1}$) = (65,47)
GeV/c$^2$ and (b) top quark production with
$m_t = 170$ GeV/c$^2$.}
\label{fig:big}
}
\end{figure}

Cut optimization was done using the {\small RGSEARCH}\cite{rgsearch}
program.  {\small RGSEARCH} uses
a modified grid search based on Monte Carlo (MC) signal events and background
samples to optimize event selection.  In this study, the MC signal samples
described above and the MC physics background samples listed in Table
\ref{tab:ind_bkg} were used.
Several combinations of selection criteria were explored starting with the
thresholds imposed by the trigger conditions.  The final selection
criteria are summarized in Table \ref{tab:datacuts}.  Other combinations
included requirements on the $E_T$\ of a second jet and/or
the azimuthal angle between the two electrons.
These combinations increased the signal to
background ratio, but reduced the signal efficiency significantly.
Values for the upper limits on $m_{ee}$ and $E_T^{\text{sum}}$\ 
were fixed while running {\small RGSEARCH}.
The cut on $m_{ee}$ was used to remove $Z \to ee$  events and that on
$E_T^{\text{sum}}$ to remove $t\overline{t} \to e e
$+jets+\mbox{${\hbox{$E$\kern-0.6em\lower-.1ex\hbox{/}}}_T$}\ events.
Distributions of $E_T^{\text{sum}}$\ for top squark production with
($m_{\widetilde t_1}$,$m_{\widetilde{\chi}^\pm_1}$) = (65,47)
GeV/c$^2$\  and Monte Carlo top quark
production are shown in Fig. \ref{fig:big}.
                                                                                
Signal detection efficiency was restricted by the reconstruction and
identification of low $E_T$\
electrons.  Only approximately 15\% of Monte Carlo events with
($m_{\widetilde t_1}$,$m_{\widetilde{\chi}^\pm_1}$) = (65,47)
GeV/c$^2$\ had two reconstructed electromagetic clusters (with an associated
track) with $E_T^{e1}>16$\ GeV and $E_T^{e2}>8$\ GeV.  In addition, the 
identification efficiency for two electrons, one with $E_T = 8$\ GeV and 
one with $E_T = 16$\ GeV, was approximately 40\%.  It is, however, essential 
to include the second electron in the selection criteria to avoid being
overwhelmed by $W$ boson events.
                                                                                
Physics backgrounds were estimated by Monte Carlo simulation or from a
combination of Monte Carlo and data.  The instrumental background from jets
misidentified as electrons was estimated entirely from data \cite{topprd}
using the jet misidentification probability
for the electron identification and kinematic cuts used in this analysis
($(6.5 \pm 1.3) \times 10^{-4}$).
Four physics backgrounds were considered in this study:
$t\overline{t}$\ production with a top quark mass of 170 GeV/c$^2$,
$WW$ production, and $Z$ boson production with final states resulting in
dielectrons.  The contribution to the background from individual
channels is given in Table \ref{tab:ind_bkg}.  The total predicted background
is $4.4 \pm 0.8$ events.
                                                                                
\begin{figure}\vbox{
\centerline{\psfig{figure=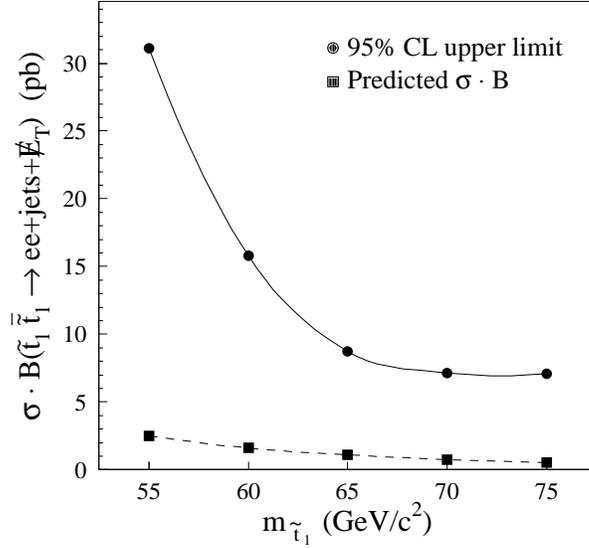,width=8cm}}
\caption{Our 95\% confidence level upper limit on $\sigma\cdot B$
as a function
of $m_{\widetilde t_1}$\ for $m_{\widetilde{\chi}^\pm_1}$ = 47
GeV/c$^2$. Also shown are the predicted values from ISAJET.}
\label{fig:limit}
}
\end{figure}

After application of the cuts to the data sample, two events remained.
Given no observed excess of events above the expected background, we set
a 95\% CL upper limit on $\sigma\cdot B$ using a Bayesian approach
\cite{bayes} with a flat prior distribution for the signal cross section.
The statistical and systematic uncertainties on the efficiency, the
integrated luminosity, and the
background estimation were included in the limit calculation with
Gaussian prior distributions.  The resulting upper limit on $\sigma\cdot B$
as a function of $m_{\widetilde t_1}$\ with fixed
$m_{\widetilde{\chi}^\pm_1}$\ = 47 GeV/c$^2$\ is shown in
Fig. \ref{fig:limit} along with the predicted $\sigma\cdot B$.
The choice of $m_{\widetilde{\chi}^\pm_1}$\ = 47 GeV/c$^2$\ allows the
widest range of $m_{\widetilde t_1}$.  As can be seen, no limit on
$m_{\widetilde t_1}$\ can be set.
The situation is similar for increased $m_{\widetilde{\chi}^\pm_1}$.
                                                                                
Although the recent results on chargino pair production from LEP limit the
likelihood for a light top squark to decay to a $b$ quark and a chargino
within the MSSM, the $\sigma\cdot B$ limit curve
shown in Fig. \ref{fig:limit} indicates the level of sensitivity in the
current D\O\ data set to a particular topology for new physics:  pair
production of new particles which decay into leptons, jets, and
non-interacting particles.  Such a new particle,
with a top-like signature, could be detectable in the current data set down
to a production cross section times branching ratio of order 10 pb.
                                                                                
We thank H. Baer and X. Tata for useful discussions.
%
We thank the staffs at Fermilab and the collaborating institutions for their
contributions to the success of this work, and acknowledge support from the 
Department of Energy and National Science Foundation (U.S.A.),  
Commissariat  \` a L'Energie Atomique (France), 
Ministries for Atomic Energy and Science and Technology Policy (Russia),
CNPq (Brazil),
Departments of Atomic Energy and Science and Education (India),
Colciencias (Colombia),
CONACyT (Mexico),
Ministry of Education and KOSEF (Korea),
CONICET and UBACyT (Argentina),
and the A.P. Sloan Foundation.
%


\begin{table}
\caption{Background contributions from individual channels.}
\begin{tabular} { c|c }
Background Channel          & Number of Events \\ \hline
$t\overline{t}$(170) $\to ee$        & $ 0.03 \pm 0.01 $ \\
$WW \to ee$                 & $ 0.02 \pm 0.01 $ \\
$Z\to ee$                   & $ 0.09 \pm 0.01 $ \\
$Z\to \tau\tau \to ee$      & $ 0.67 \pm 0.13 $ \\
Misidentification           & $ 3.6 \pm 0.8  $ \\ \hline
Total                       & $ 4.4 \pm 0.8 $ \\
\end{tabular}
\label{tab:ind_bkg}
\end{table}
                                                                                
\begin{table}
\caption{Kinematic cuts. $E_T^{\text{sum}}$\ is defined in the text.}
\begin{tabular} { r@{~}l@{~}r@{~}l }
$E^{e1}_{T}$ & $\geq$ & 16 & GeV    \\
$E^{e2}_{T}$ & $\geq$ &  8 & GeV    \\
$E^{j1}_{T}$ & $\geq$ & 30 & GeV    \\
\mbox{${\hbox{$E$\kern-0.6em\lower-.1ex\hbox{/}}}_T$} & $\geq$ & 22 & GeV    \\
$m_{ee}$     & $\leq$ & 60 & GeV/c$^2$ \\
$E_T^{\text{sum}}$    & $\leq$ & 90 & GeV    \\
\end{tabular}
\label{tab:datacuts}
\end{table}
                                                                                

\end{document}